\documentclass[]{article}
\usepackage{amssymb}
\usepackage{algorithmic}
\usepackage{hyperref}
\usepackage[toc,page]{appendix}
\usepackage{listings}
\usepackage{graphicx}
\usepackage[hang,flushmargin]{footmisc} 

\title{XCRUSH: A Family of ARX Block Ciphers}
\author{
	Evan Saulpaugh\\
	\href{mailto:evan.saulpaugh@gmail.com}{evan.saulpaugh@gmail.com}
}
\date{September 8, 2015}

\newcommand\blfootnote[1]{%
	\begingroup
	\renewcommand\thefootnote{}\footnote{#1}%
	\addtocounter{footnote}{-1}%
	\endgroup
}

\begin{document}
\maketitle

\hfill\\
\textbf{Abstract.} {The XCRUSH family of non-Feistel, ARX block ciphers is designed to make efficient use of modern 64-bit general-purpose processors using a small number of encryption rounds which are simple to implement in software. The avalanche function, which applies one data-dependent, key-dependent rotation per 64-bit word of plaintext per round, allows XCRUSH to produce an almost totally diffuse 256-bit block after just the first two rounds. Designed for speed in software, the reference implementation for 3-round XCRUSH is measured at $\sim$7.3 cycles/byte single-threaded on an Intel Haswell processor. A deterministic random bit generator, constructed using the avalanche function, serves as a key scheduling algorithm. No security claims are made in this paper.}\\

\noindent \textbf{Keywords: }ARX, data-dependent rotations, DDR, XCRUSH
\blfootnote{The algorithms presented in this paper are free from any intellectual property restrictions. This release does not constitute an endorsement of these algorithms for official use.}

\section{Introduction}
This author was originally inspired to create a new block cipher more efficient than Rijndael\cite{aes} in software run on modern general-purpose processors. The idea was to use 64-bit XOR, add, and rotate operations instead of lookup tables which work with only eight or sometimes 16 bits at a time, thinking that perhaps the technique was without precedent. Had it been, this cipher would likely have been called \textit{Czar} or \textit{Tsar}, a play on the letters XAR, meaning XOR, Add, Rotate.\\
\\
In experimenting with rotations of a 256-bit block using eight bits of key to determine the rotation distance, it became obvious that key-dependent rotations were just as ineffective as fixed rotations for achieving diffusion quickly. Subsequently and by default, data-dependent rotations came to be the tool of choice.\\
\\
XCRUSH has a fixed block length of 256 bits and supports key sizes of 128, 192, and 256 bits. Developed uninfluenced by any prior work involving data-dependent rotations, the techniques presented in this paper are the result of this author's attempts to achieve fast diffusion in software with 64-bit arithmetic.\\

\begin{table}[h!]
	\begin{center}
		\textbf{Table 1. }{Notation}\\
		\hfill \\
		\begin{tabular}{cl}
			Notation & Description\\
			\hline
			$x + y$ & addition of $x$ and $y$ modulo $2^n$\\
			$x * y$ & multiplication of $x$ and $y$ modulo $2^n$\\
			$x \oplus y$ & bitwise XOR of $x$ and $y$\\
			$x\land y$ & bitwise AND of $x$ and $y$\\
			$x\gg i$ & right shift of $x$ by $i$ bits\\
			$x\lll i$ & left rotation of $x$ by $i$ bits\\
			\hline
		\end{tabular}
	\end{center}
\end{table}

\section{Description}
\subsection[]{Algorithm}
Let $P = (P_1,P_2,P_3,P_4)$ and $C = (C_1,C_2,C_3,C_4)$ be the 256-bit plaintext and ciphertext respectively where $P_i$ and $C_i$ are 64-bit words. Let $SK_j$ be the 64-bit subkey for $j = 1,...,16$. Then, the encryption process can be formulated as in Algorithm 1.
\\
\rule{\linewidth}{0.8pt}\\
\textbf{Algorithm 1 }{The Encryption Process of XCRUSH}\\
\rule{\linewidth}{0.4pt}
\begin{algorithmic}
	\STATE $(T_1,T_2,T_3,T_4) = P = (P_1,P_2,P_3,P_4)$
	\STATE $SK = (SK_1,SK_2,...,SK_{16})$
	\FOR {$i = 0$ to $2$}
	\STATE $j = i * 4$
	\STATE $(RK_1,RK_2,RK_3,RK_4) = (SK_{j+1},SK_{j+2},SK_{j+3},SK_{j+4})$
	\STATE $T_1 = A(T_1, T_2 + T_3 + T_4 + RK_1)$
	\STATE $T_2 = A(T_2, T_1 + T_3 + T_4 + RK_2)$
	\STATE $T_3 = A(T_3, T_1 + T_2 + T_4 + RK_3)$
	\STATE $T_4 = A(T_4, T_1 + T_2 + T_3 + RK_4)$
	\ENDFOR
	\STATE $(C_1,C_2,C_3,C_4) = (T_1 \oplus SK_{13},T_2 \oplus SK_{14},T_3 \oplus SK_{15},T_4 \oplus SK_{16})$
\end{algorithmic}
\rule{\linewidth}{0.4pt}
\\
\\
\textbf {\textit{A}-Function } {{\textit{A}-Function is a keyed avalanche function that provides both confusion and diffusion. $C$-Function returns an integer on interval [0,63] which is used as the bitwise leftward rotation distance. $A$-Function is used in the encryption process and as part of the key expansion. Where $X$ and $A$ are 64-bit words, \textit{A}-Function can be formulated as in Algorithm 2.}
\\
\rule{\linewidth}{0.8pt}\\
\textbf{Algorithm 2 }{{\textit{A}-Function}\\
\rule{\linewidth}{0.4pt}\\
$A(X, A) = (Y)$\\
\indent $Y = (X + A) \lll C(A)$\\
\rule{\linewidth}{0.4pt}\\
\\
\textbf {\textit{C}-Function } {{\textit{C}-Function is a chaotic compression designed to produce a large set of 6-bit values when applied to a set of very similar inputs. Where $X$ is a 64-bit word, \textit{C}-Function can be formulated as in Algorithm 3.}
\\
\rule{\linewidth}{0.8pt}\\
\textbf{Algorithm 3 }{{\textit{C}-Function}\\
\rule{\linewidth}{0.4pt}\\
$C(X) = (Y)$\\
\indent $X = (X \gg 32) + X$\\
\indent $X = (X \gg 11) \oplus X$\\
\indent $X = (X \gg 9) + X$\\
\indent $X = (X \gg 6) + X$\\
\indent $Y = X \land 0x3f$\\
\rule{\linewidth}{0.4pt}\\
\subsection[]{Key Expansion}
The 16 64-bit subkeys necessary for the encryption process are generated by seeding a pseudorandom number generator with the encryption key. Let the encryption key $K$ be equal to $(K_1,K_2)$, $(K_1,K_2,K_3)$, and $(K_1,K_2,K_3,K_4)$ for key lengths 128, 192, and 256 bits respectively, where each $K_i$ is a 64-bit word. Let the 320-bit PRNG seed $S = (S_1,S_2,S_3,S_4,S_5)$ be equal to $(K_1,K_2,C,C,C)$, $(K_1,K_2,K_3,C,C)$, and $(K_1,K_2,K_3,K_4,C)$ for key lengths 128, 192, and 256 bits respectively, where $C = 4142135623730950488$ and $S_i$, $K_i$, and $C$ are 64-bit words. Then, the key expansion can be formulated as in Algorithm 4.
\\
\rule{\linewidth}{0.8pt}\\
\textbf{Algorithm 4 }{The Key Expansion Algorithm}\\
\rule{\linewidth}{0.4pt}
\begin{algorithmic}
	\FOR{$i = 1$ to 10} 
	\STATE $(Y,S) = N(S)$
	\ENDFOR
	\STATE $SK = (SK_1,SK_2,...,SK_{16})$
	\FOR{$i = 1$ to 16}
	\STATE $(Y,S) = N(S)$
	\STATE $SK_i = Y$
	\ENDFOR
\end{algorithmic}
\rule{\linewidth}{0.4pt}\\
\rule{\linewidth}{0.8pt}\\
\textbf{Algorithm 5 }{{\textit{N}-Function is a pseudorandom number generator with 320 bits of state. \textit{N}-Function uses the avalanche function to generate a series of 64-bit outputs.}\\
\rule{\linewidth}{0.4pt}\\
$N(S) = (Y,S)$\\
\indent $S = (S_1,S_2,S_3,S_4,S_5)$\\
\indent $T = S_2$\\
\indent $S_2 = S_3$\\
\indent $S_3 = S_4$\\
\indent $S_4 = S_5$\\
\indent $S_5 = S_1$\\
\indent $Y = S_1 = A(S_1, S_1 + T)$\\
\indent $S = (S_1,S_2,S_3,S_4,S_5)$\\
\rule{\linewidth}{0.4pt}\\
\section[]{Design Rationale}

\subsection[]{Algorithm}

By making the addition to, but more importantly the bitwise rotation of each 64-bit word of plaintext dependent on the other three words in the plaintext block, and by allowing the output of one avalanche operation to feed into the next, we can ensure that a small change in the plaintext block will tend to result in a very large change in the output block after each round.\\
\\
Given this construction, it appeared natural to simply add a 64-bit word of key material to the sum of the three words of plaintext going into the avalanche function. In this way, confusion and diffusion are achieved efficiently and simultaneously.\\
\\
Finally, as key whitening, the block is XORed with the final 256 bits of subkey in order to obscure all the inputs into the final invokations of the avalanche function.

\subsection[]{Key Expansion}

The key expansion is accomplished by seeding a deterministic random bit generator (PRNG) with the encryption key to produce the subkeys used in the encryption process. The PRNG used is based on xorshift generators\cite{marsaglia03} and is chosen for speed and simplicity of implementation.\\
\\
The complex one-way nature of the key expansion is intended to make it difficult for an attacker to use information about a subkey to recover infomation about the other subkeys or about the master key. Additionally, the PRNG method of subkey derivation, by which a stream of subkeys is dependent on all the bits of the master key, is intended to mitigate related-key attacks.\\
\\
The constant $C$, namely the leading decimal digits of the fractional part of the square root of two, is arbitrarily chosen to ensure that the PRNG is not seeded with all zeros.\\
\\
Note that, regardless of key size, XCRUSH is composed of three avalanche rounds and a key-whitening XOR.

\begin{figure}[p]
	\centering
	\includegraphics[width=6in]{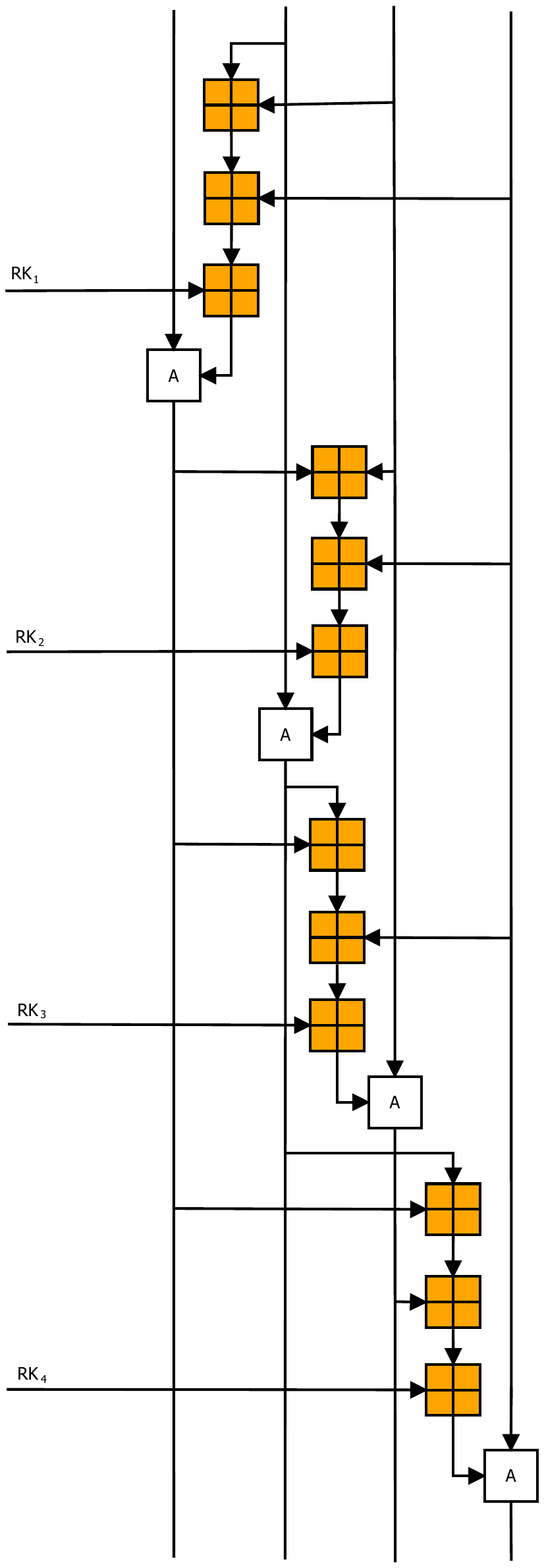}
	\caption[]
	{A round of XCRUSH.}
\end{figure}
\begin{figure}[p]
	\centering
	\includegraphics[height=6in]{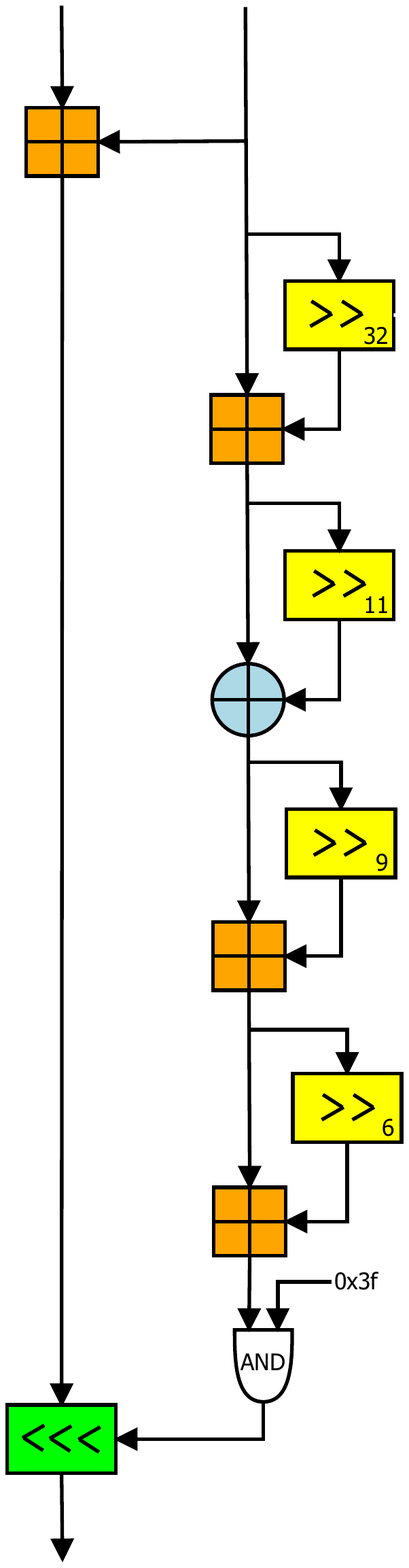}
	\caption[]
	{The avalanche function.}
\end{figure}

\section{Concluding Remarks}

XCRUSH is an ambitious attempt at creating an unconventional family of block ciphers which may or may not lend itself well to traditional cryptanalysis. It is the hope of this author that it may serve to advance the field of cryptography in its own way.

\newpage
\begin{appendices}
\section{Test Vectors}
\subsection{XCRUSH-128}
\begin{lstlisting}
KEY:
    1599D14129204267 E4C91210F1C15541
PLAINTEXT:
    9338192346089EEE 965D12810033DDF0 434C5669E9E31202 86416B3296055DC1
CIPHERTEXT:
    2AC5C0D9B62355A2 9DEFB4F22A3D6DBF CC18261B50072FBC CCB953C4947A6C39
\end{lstlisting}
\subsection{XCRUSH-192}
\begin{lstlisting}
KEY:
    4211121041C35A31 E4E4961BB81941BA CC982462195662AA
PLAINTEXT:
    4440306090522AB0 31249688284691DF 4C15654900DB1A19 19A0FF64135229D2
CIPHERTEXT:
    2FEFD41974AFDD44 15BA6339E5C03563 42BA28CF31B5F400 CCD58FC905686D9F
\end{lstlisting}
\subsection{XCRUSH-256}
\begin{lstlisting}
KEY:
    F0E0D0C0B0A09080 7060504030201000 F1D3B597795B3D1F 021346578A9BCEDF
PLAINTEXT:
    311D411620304361 48165C7790022614 9536295B87012640 396218842A490866
CIPHERTEXT:
    000947604A76E469 E34346B03745CAC9 244D96ACC783C42B 95406757BE5653D9
 
\end{lstlisting}
\newpage
\section{Reference Implementation}
\begin{verbatim}
/*
The C reference implementation of the
block cipher family XCRUSH.

This software is released into the public domain.
*/

#include <stdio.h>

#define BLOCK_LENGTH_LONGS 4
#define KEY_LEN_128_BITS 2
#define KEY_LEN_192_BITS 3
#define KEY_LEN_256_BITS 4
#define NUM_SUBKEYS 16
#define C 4142135623730950488L
#define LONG_LONG_SIZE 64

int compress(unsigned long long x) {
    x = (x >> 32) + x;
    x = (x >> 11) ^ x;
    x = (x >> 9) + x;
    return ((x >> 6) + x) & 0x3f;
}

long long avalanche(unsigned long long v, unsigned long long a) {
    v += a;
    const int shiftAmount = compress(a);
    /* ROTATE LEFT */
    return (v << shiftAmount) | (v >> (LONG_LONG_SIZE - shiftAmount));
}

long long unavalanche(unsigned long long v, unsigned long long a) {
    const int shiftAmount = compress(a);
    /* ROTATE RIGHT */
    v = (v >> shiftAmount) | (v << (LONG_LONG_SIZE - shiftAmount));
    return v - a;
}

void _encrypt(unsigned long long plaintext[], int offset, int length,
        unsigned long long subkeys[16]) {

    const long __0 = subkeys[0];
    const long __1 = subkeys[1];
    const long __2 = subkeys[2];
    const long __3 = subkeys[3];
    const long __4 = subkeys[4];
    const long __5 = subkeys[5];
    const long __6 = subkeys[6];
    const long __7 = subkeys[7];
    const long __8 = subkeys[8];
    const long __9 = subkeys[9];
    const long _10 = subkeys[10];
    const long _11 = subkeys[11];

    const long _12 = subkeys[12];
    const long _13 = subkeys[13];
    const long _14 = subkeys[14];
    const long _15 = subkeys[15];

    const int end = offset + length;

    int one___, two___, three_;

    long a_, b_, c_, d_, temp;

    // for each block
    for ( ; offset < end; offset += BLOCK_LENGTH_LONGS) {

        one___ = offset + 1;
        two___ = offset + 2;
        three_ = offset + 3;

        a_ = plaintext[offset];
        b_ = plaintext[one___];
        c_ = plaintext[two___];
        d_ = plaintext[three_];

        /* round 1 */
        temp = c_ + d_;
        a_ = avalanche(a_, temp + b_ + __0);
        b_ = avalanche(b_, temp + a_ + __1);
        temp = a_ + b_;
        c_ = avalanche(c_, temp + d_ + __2);
        d_ = avalanche(d_, temp + c_ + __3);

        /* round 2 */
        temp = c_ + d_;
        a_ = avalanche(a_, temp + b_ + __4);
        b_ = avalanche(b_, temp + a_ + __5);
        temp = a_ + b_;
        c_ = avalanche(c_, temp + d_ + __6);
        d_ = avalanche(d_, temp + c_ + __7);

        /* round 3 */
        temp = c_ + d_;
        a_ = avalanche(a_, temp + b_ + __8);
        b_ = avalanche(b_, temp + a_ + __9);
        temp = a_ + b_;
        c_ = avalanche(c_, temp + d_ + _10);
        d_ = avalanche(d_, temp + c_ + _11);

        plaintext[offset] = a_ ^ _12;
        plaintext[one___] = b_ ^ _13;
        plaintext[two___] = c_ ^ _14;
        plaintext[three_] = d_ ^ _15;
    }
}

void decrypt(unsigned long long ciphertext[], int offset,
        int length, unsigned long long subkeys[16]) {

    const long __0 = subkeys[0];
    const long __1 = subkeys[1];
    const long __2 = subkeys[2];
    const long __3 = subkeys[3];
    const long __4 = subkeys[4];
    const long __5 = subkeys[5];
    const long __6 = subkeys[6];
    const long __7 = subkeys[7];
    const long __8 = subkeys[8];
    const long __9 = subkeys[9];
    const long _10 = subkeys[10];
    const long _11 = subkeys[11];

    const long _12 = subkeys[12];
    const long _13 = subkeys[13];
    const long _14 = subkeys[14];
    const long _15 = subkeys[15];

    const int end = offset + length;

    int one___, two___, three_;

    long a_, b_, c_, d_, temp;

    for ( ; offset < end; offset += BLOCK_LENGTH_LONGS) {

        one___ = offset + 1;
        two___ = offset + 2;
        three_ = offset + 3;

        a_ = ciphertext[offset] ^ _12;
        b_ = ciphertext[one___] ^ _13;
        c_ = ciphertext[two___] ^ _14;
        d_ = ciphertext[three_] ^ _15;

        temp = a_ + b_;
        d_ = unavalanche(d_, temp + c_ + _11);
        c_ = unavalanche(c_, temp + d_ + _10);
        temp = c_ + d_;
        b_ = unavalanche(b_, temp + a_ + __9);
        a_ = unavalanche(a_, temp + b_ + __8);

        temp = a_ + b_;
        d_ = unavalanche(d_, temp + c_ + __7);
        c_ = unavalanche(c_, temp + d_ + __6);
        temp = c_ + d_;
        b_ = unavalanche(b_, temp + a_ + __5);
        a_ = unavalanche(a_, temp + b_ + __4);

        temp = a_ + b_;
        d_ = unavalanche(d_, temp + c_ + __3);
        c_ = unavalanche(c_, temp + d_ + __2);
        temp = c_ + d_;
        b_ = unavalanche(b_, temp + a_ + __1);
        a_ = unavalanche(a_, temp + b_ + __0);

        ciphertext[offset] = a_;
        ciphertext[one___] = b_;
        ciphertext[two___] = c_;
        ciphertext[three_] = d_;
    }
}

long long S_1,  S_2,  S_3,  S_4,  S_5;

long long next() {
    long long t = S_2;
    S_2 = S_3;
    S_3 = S_4;
    S_4 = S_5;
    S_5 = S_1;
    S_1 = avalanche(S_1, S_1 + t);
    return S_1;
}

void expand_key(unsigned long long key[], int keyLen,
        unsigned long long subkeybuf[NUM_SUBKEYS]) {
    switch (keyLen) {
    case KEY_LEN_128_BITS:
        S_1 = key[0];
        S_2 = key[1];
        S_3 = C;
        S_4 = C;
        S_5 = C;
        break;
    case KEY_LEN_192_BITS:
        S_1 = key[0];
        S_2 = key[1];
        S_3 = key[2];
        S_4 = C;
        S_5 = C;
        break;
    case KEY_LEN_256_BITS:
        S_1 = key[0];
        S_2 = key[1];
        S_3 = key[2];
        S_4 = key[3];
        S_5 = C;
        break;
    default:
        break;
    }

    for(int i = 0; i < 10; i++) {
        next();
    }
    for(int i = 0; i < NUM_SUBKEYS; i++) {
        subkeybuf[i] = next();
    }
}

int main(int argc, const char * argv[]) {

    const int keyLen = KEY_LEN_256_BITS;

    unsigned long long key[keyLen] = {
        0xF0E0D0C0B0A09080L,
        0x7060504030201000L,
        0xF1D3B597795B3D1FL,
        0x021346578A9BCEDFL
    };

    const int dataLen = 4;
    unsigned long long data[dataLen] = {
        0x311d411620304361L,
        0x48165c7790022614L,
        0x9536295b87012640L,
        0x396218842a490866L
    };

    unsigned long long subkeys[NUM_SUBKEYS];
    expand_key(key, keyLen, subkeys);

    for(int i = 0; i < dataLen; i++) {
        printf("%#llx ", data[i]);
    }

    printf("\n");

    _encrypt(data, 0, dataLen, subkeys);

    for(int i = 0; i < dataLen; i++) {
        printf("%#llx ", data[i]);
    }

    printf("\n");

    decrypt(data, 0, dataLen, subkeys);

    for(int i = 0; i < dataLen; i++) {
        printf("%#llx ", data[i]);
    };

    return 0;
}

\end{verbatim}

\end{appendices}

\end{document}